\documentclass[twocolumn,showpacs,prl,amsmath,amssymb]{revtex4-1}
\usepackage{graphicx}
\usepackage{bm}
\usepackage{color}
\usepackage{subfigure}
\DeclareMathOperator{\Tr}{Tr} 

\DeclareMathOperator{\Max}{Max}

\begin{document}

\title{Quantum read-out and fast initialization of nuclear spin qubits with electric currents}
\author{Noah Stemeroff}
\author{Rog\'{e}rio de Sousa}
\email[]{rdesousa@uvic.ca}
\affiliation{Department of Physics and Astronomy, University of Victoria,
Victoria, B.C., V8W 3P6, Canada}
\date{\today}

\begin{abstract}
  Nuclear spin qubits have the longest coherence times in the solid
  state, but their quantum read-out and initialization is a great
  challenge.  We present a theory for the interaction of an electric
  current with the nuclear spins of donor impurities in
  semiconductors. The theory yields a sensitivity criterion for
  quantum detection of nuclear spin states using electrically detected
  magnetic resonance, as well as an all electrical method for fast
  nuclear spin qubit initialization.
\end{abstract}


\pacs{
03.65.Yz; 
03.67.Lx; 
76.70.-r. 
}

\maketitle



Running a current over a localized electron spin provides an extremely
sensitive method to detect electron spin resonance \cite{honig66}.
This technique, commonly referred to as ``electrically detected
magnetic resonance'' (EDMR) is now experiencing a revival in the
context of solid state quantum computing \cite{kane98}. Several interesting recent
results
\cite{boehme03,mccamey06,stegner07,sarovar08,morley08,desousa09,morishita09,mccamey10,hoehne11,lo11}
point to the possibility of using EDMR for quantum read-out of
individual donor nuclear spin states in silicon. The hope is that EDMR
will be the ``silicon alternative'' to optically detected magnetic
resonance, a method that can detect single nuclear spin states of
defects in diamond \cite{jelezko04}. Nuclear spins in silicon are
exceptional quantum memory units, with coherence times $T_{2n}$
over a few seconds \cite{ladd05,morton08}. While single spin
detection of donor electrons in silicon was demonstrated recently
\cite{morello10}, the corresponding read-out of individual nuclear
spins remains a great open problem.

Nuclear spin read-out using EDMR \cite{sarovar08} is based on the
interaction between donor electron ($\bm{S}$) and nuclear ($\bm{I}$) spin
in an external magnetic field $B$:
\begin{equation}
{\cal H}_{en}= \omega_e  S_{z} - \omega_n  I_{z} + A\bm{S}\cdot \bm{I},
\label{hyp}
\end{equation}
where $\omega_e=g_e \mu_e B/\hbar$ and $\omega_n=g_n\mu_n B/\hbar$ are
the Zeeman frequencies for electron and nuclear spins, respectively.
The hyperfine interaction $A$ couples the donor electron to its
nuclear spin. From the approximation ${\cal H}_{en}\approx (\omega_e +
A I_z)S_z$, we see that the electron spin resonance frequency will
depend on the nuclear spin state.  The EDMR experiment consists in
applying microwaves at a fixed frequency $\omega_{\rm{ESR}}$, and
measuring an electrical current as a function of the applied magnetic
field $B$.  Thus, the current will show a single peak either at
$B_{-}=\hbar(\omega_{\rm{ESR}}-A/2)/(g_e\mu_e)$ when the donor nuclear
spin is aligned along the $B$ field, or at
$B_{+}=\hbar(\omega_{\rm{ESR}}+A/2)/(g_e\mu_e)$ when the donor nucleus is
aligned against the field.  It is therefore believed that the
detection of electron spin resonance of a single donor impurity using
ultra sensitive EDMR methods will allow the read-out of its nuclear
spin state, or even of $^{29}$Si nuclear spins nearby the donor.

However, it is not clear that this is possible. The problem is that
the time it takes to detect EDMR of a single donor spin
$t_{\rm{EDMR}}$ must be \emph{shorter} than nuclear spin-flip time
$T_{1n}$, and as we show below the electric current will induce
nuclear spin-flips.  Moreover, read-out time must be \emph{longer}
than nuclear spin coherence time $T_{2n}$ for the nuclear spin wave
function to collapse into one of the outcome states.  Thus, the EDMR
read-out scheme will be limited by a contrast of
$\exp{[-\Max{(T_{2n},t_{\rm{EDMR}})}/T_{1n}]}$.  The best possible
scenario occurs when $t_{\rm{EDMR}}< T_{2n}$, leading to an optimal
read-out contrast of $\exp{(-T_{2n}/T_{1n})}$.

In this letter, we describe our theory for the interaction of nuclear
spin qubits with a ``classical current'', i.e. a current formed by an
incoherent ensemble of conduction electrons, the one used in common
electronic devices such as EDMR.  

Consider a donor impurity interacting with an electron
gas \cite{boehme03,mccamey06,stegner07,sarovar08,morley08,desousa09,morishita09,mccamey10,hoehne11,lo11}.
Scattering events preserve the total spin of conduction and localized
electrons, leading to the interaction Hamiltonian \cite{hewson93}
\begin{eqnarray}
{\cal H}_{\rm{ce}}&=& J\sum_{k,k'} \left[S_{-}c^{\dag}_{k'\uparrow}c_{k\downarrow} +S_{+}c^{\dag}_{k'\downarrow}c_{k\uparrow}\right.\nonumber\\
&&\left.+S_{z}\left(c^{\dag}_{k'\uparrow}c_{k\uparrow}-c^{\dag}_{k'\downarrow}c_{k\downarrow}\right)
\right].
\label{hce}
\end{eqnarray}
Here $c^{\dag}_{k\sigma},c_{k\sigma}$ are creation/annihilation operators for a conduction electron
with momentum $k$ and spin $\sigma=\uparrow$ or $\downarrow$,
$S_{\pm}=S_{x}\pm i S_{y}$ are spin raising/lowering operators for the
donor electron, and $J$ is their ``bare'' exchange interaction.

\begin{figure}
\includegraphics[width=0.5\textwidth]{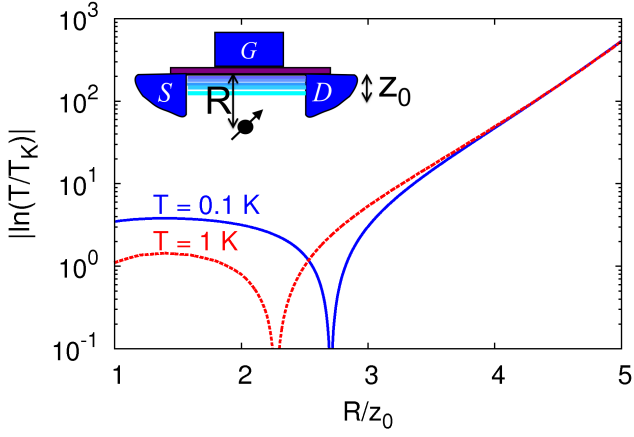}
\caption{(color online) This figure demonstrates how a donor impurity
  implanted in a silicon accumulation field effect transistor (inset)
  can be tuned into and out of the Kondo regime using a top gate
  voltage. The plot shows $|\ln{(T/T_K)}|$ for $T=0.1$~K and $T=1$~K
  for a single donor impurity fixed at a distance $R$ from the
  transistor interface (model assumptions and parameters are the same
  as Fig.~7 of \cite{desousa09}).  The length scale $z_0$ models the
  two-dimensional electron gas width, that is controllable
  electrically over the range $z_0\approx 30-100$~\AA. Hence a donor
  can be tuned from $R/z_0\approx 3$ (strong coupling regime) to
  $R/z_0 \approx 5$ (weak coupling regime) by increasing the top gate
  voltage. Recently, a similar electrical tuning of the donor Kondo
  temperature was demonstrated in a side-gated silicon device
  \cite{lansbergen10}.}
\label{fig1}
\end{figure}

Exchange scattering will tend to equilibrate the localized
spin with the electron gas. If the localized spin is not in thermal
equilibrium, its expectation value 
$\langle S_{\alpha}\rangle$ 
along any direction 
$\alpha=x,y,z$ will decay exponentially in time according to the rate,
\begin{equation}
\Gamma_e\equiv \frac{1}{T_{1e}} = \frac{1}{T_{2e}}=\frac{2\pi}{\hbar}|J_{\rm{eff}}\nu|^{2} 
\hbar\omega_e \coth{\left(\frac{\hbar\omega_e}{2 k_B T}\right)},
\label{gammae}
\end{equation}
where $\nu$ is the energy density of states at the Fermi level. The effective exchange interaction
$J_{\rm{eff}}$ is obtained after summing and renormalizing all orders of perturbation
theory in $J\nu$, resulting in an universal expression that depends only on the Kondo temperature
$T_K$ \cite{hewson93,vanhaesendonck87}:
\begin{equation}
\left|J_{\rm{eff}}\nu\right|^{2}= \left[ \pi^2 + \frac{4}{3}\left| \ln{\left(\frac{T}{T_K}\right)}\right|^{2}\right]^{-1}.
\label{jeff}
\end{equation}
Calculations of the Kondo temperature for realistic
parameters of a silicon transistor are presented in Figure~\ref{fig1}.

Equation (\ref{gammae}) is valid for $T>T_K$. When $T<T_K$, the
conduction electron spins will screen out the donor electron spin,
forming a Kondo singlet. This will add non-exponential decay to
$\langle S_\alpha\rangle$, with a rate $\Gamma_e\sim k_B T_K /\hbar$
\cite{lobaskin05}.

At $T>T_K$, $\langle S_{\alpha}\rangle$ decays exponentially in time,
towards its thermal equilibrium value.  This leads to the following
effective model for donor electron plus nuclear spin evolution subject
to the conduction electron environment \cite{rikitake05}:
\begin{equation}
\frac{\partial \hat{\rho}}{\partial t} = 
-i\left[{\cal H}_{en},\hat{\rho}\right] 
+\Gamma_e \left(\sum_{\alpha=x,y,z} S_{\alpha}\hat{\rho}S_{\alpha}-\frac{3}{4}\hat{\rho}+ p_e S_{z}\right).
\label{liouville}
\end{equation}
Here $\hat{\rho}$ is the density matrix describing donor electron plus
nuclear spin, $\Gamma_e$ is given by Eq.~(\ref{gammae}), and
$p_e=-\tanh{[\hbar\omega_e/(2k_B T)]}$ is the equilibrium donor
electron spin polarization. With the usual definitions
$\langle\bm{S}\rangle=\Tr\{\hat{\rho}\bm{S}\}$,
$\langle\bm{I}\rangle=\Tr\{\hat{\rho}\bm{I}\}$, and
$\langle\bm{S}\bm{I}\rangle =\Tr{(\hat{\rho}\bm{SI})}$ (a matrix
formed by the outer product between $\bm{S}$ and $\bm{I}$) we can put
Eq.~(\ref{liouville}) in a more convenient form that we call the
generalized Bloch equation for the donor:
\begin{subequations}
\begin{eqnarray}
\dot{\left\langle \bm{S}\right\rangle} &=&\omega_e \hat{\bm{z}}\times \left\langle \bm{S}\right\rangle 
-A \left\langle \bm{S}\times \bm{I} \right\rangle -\Gamma_e \left(\left\langle \bm{S}\right\rangle 
- \frac{p_e}{2}\hat{\bm{z}}\right),\label{dots}\\
\dot{\left\langle \bm{I}\right\rangle} &=&-\omega_n \hat{\bm{z}}\times \left\langle \bm{I}\right\rangle 
+A \left\langle \bm{S}\times \bm{I} \right\rangle ,\label{doti}\\
\dot{\left\langle \bm{S}\bm{I}\right\rangle} &=& \omega_e \hat{\bm{z}}\times \left\langle \bm{S}\bm{I}\right\rangle
+\omega_n \left\langle \bm{S}\bm{I}\right\rangle \times \hat{\bm{z}}\nonumber\\ 
&&+\frac{A}{4} \left(\bm{S}-\bm{I}\right)\cdot \bm{\varepsilon}
-\Gamma_e \left\langle \bm{S}\bm{I}\right\rangle,\label{dotsi}
\end{eqnarray}
\end{subequations} 
with
$\bm{\varepsilon}=\varepsilon_{\alpha\beta\gamma}\bm{\hat{e}}_{\alpha}\bm{\hat{e}}_{\beta}\bm{\hat{e}}_{\gamma}$
the Levi-Civita tensor.

We can solve Equations~(\ref{dots})--(\ref{dotsi}) analytically by
moving to a reference frame where the electron spin rotates with
frequency $\omega_e$ and the nuclear spin rotates with
$-\omega_n$; within order $(A/\omega_e)^{2}$ we get 
\begin{subequations}
\begin{eqnarray}
  \dot{\left|\left\langle \bm{I'}_{\perp}\right\rangle\right|} 
&\approx& -\frac{1}{2}\frac{\Gamma_e}{1+2\left(\frac{\Gamma_e}{A}\right)^{2}}
  \left|\left\langle \bm{I'}_{\perp}\right\rangle\right|,\label{iperp}\\
  \dot{\langle I_{z}\rangle}&\approx&-\frac{\Gamma_e}{1+2\left(\frac{\Gamma_e}{A}\right)^{2}
+2\left(\frac{\tilde{B}}{A}\right)^{2}}
  \left(\langle I_{z}\rangle - \frac{p_e}{2}\right),\label{iz}
\end{eqnarray}
\end{subequations}
where $\left\langle \bm{I'}\right\rangle$ is the nuclear spin operator
in the rotating frame, and $\tilde{B}=(\omega_e + \omega_n)$. 
The decay rate of Eq.~(\ref{iperp}) is the new nuclear spin decoherence rate induced by the current,
\begin{equation}
\frac{1}{T_{2n}}= \frac{1}{2}\frac{\Gamma_e}{1+2\left(\frac{\Gamma_e}{A}\right)^{2}}.
\label{1t2n}
\end{equation}
In the limit $\Gamma_e \ll A$, we get
$1/T_{2n}\approx \Gamma_e/2$ showing that the nuclear spin
``follows'' the electron.  However, in the opposite regime of
$\Gamma_e\gg A$ we get a decoherence rate that is inversely
proportional to $\Gamma_e$: $1/T_{2n}\approx
A^{2}/\Gamma_e$, signaling a transition to a motional
narrowing regime.  Here the electron spin fluctuates so fast that its
effect on the nuclear spin's phase is averaged out.

Similarly, the decay rate of Eq.~(\ref{iz}) is the new nuclear spin-flip rate,
\begin{equation}
\frac{1}{T_{1n}}= \frac{\Gamma_e}{1+2\left(\frac{\Gamma_e}{A}\right)^{2}
+2\left(\frac{\tilde{B}}{A}\right)^{2}}.
\label{1t1n}
\end{equation}
As $\Gamma_e$ increases, $1/T_{1n}$ peaks at
$\Gamma_e=\tilde{B}\approx \omega_e$.  The origin of this effect is a
transfer of population from the electron spin to the nucleus assisted
by the ``flip-flop'' coherence $\left\langle S_{-}I_{+}\right\rangle$.
It is necessary for the electron spin fluctuation frequency $\Gamma_e$
to match the flip-flop coherence $\tilde{B}$ for the transfer to
occur. The net result is a transfer of electron spin polarization
$p_e$ into the nuclear spin.

This result suggests a method for qubit initialization via interaction
with the electron gas.  The idea is to electrically tune the Kondo
temperature so that $\Gamma_e$ gets close to $\omega_e$. This can be
achieved when $\ln{(T/T_K)}\approx \sqrt{3\pi/(2|p_e|)}$ or $T\approx 10
\;T_K$ for $|p_e|=1$, and we call this the ``strong coupling regime''
($\Gamma_e\approx \omega_e$).  Figure~\ref{fig1} describes how to tune
into this regime with a top gated device. In the strong coupling
regime we get $\Max{(1/T_{1n})}\approx A^{2}/(4\omega_e)$.  For a
phosphorus donor in silicon, $A=120\;\rm{MHz}$ and $\omega_e = 28$~GHz
for $B=1$~T, leading to qubit initialization times $T_{1n}=1$~$\mu$s.
These electrical initialization times are several orders of magnitude
faster than other methods previously considered for donor nuclear
spins in silicon.  In \cite{mccamey09}, white light was used to
polarize donor nuclei within $150$~s, while \cite{yang09} used a laser
tuned to an exciton transition to initialize within $0.5$~s.

We remark that the nuclear spin-flip process described by
Eq.~(\ref{iz}) results from an ``indirect'' interaction of the donor
nucleus with conduction electrons, mediated by the donor electron spin
through hyperfine coupling $A$.  There also exists an additional
``direct'' interaction between the donor nuclear spin and the
conduction electrons: It gives rise to the Korringa relaxation
mechanism \cite{abragam61}, that is usually quite weak for
semiconductors. We calculated the Korringa rate for the phosphorus
donor in the silicon transistor shown in Fig.~\ref{fig1}. We obtained
$(1/T_{1n})_{\rm{Korringa}}< 10^{-3}$~s$^{-1}$ for $R/z_0>2$ at low
temperature ($T<10$~K). Hence, at times shorter than a few minutes we
may neglect this ``direct'' interaction between the donor nucleus and
the electron gas, and the donor nuclear spin polarization will remain
close to the electron's, $p_n\approx
p_e=-\tanh{(\hbar\omega_e/2k_BT)}$. However, we note that the
Korringa rate will eventually drive the nuclear spin back to its
thermal equilibrium state, with small polarization $p_n= +\tanh{(\hbar
  \omega_n/2k_BT)}$.

We now describe nuclear spin read-out with EDMR. The donor spin is
irradiated with a resonant microwave field of frequency
$\omega_{\rm{ESR}}$ and amplitude $B_{\perp}$, inducing electron spin
precession with frequency $\Omega_{\perp}=g_e\mu_e B_{\perp}/\hbar$.
Only electron spins satisfying the equal population or ``saturation''
condition $\Omega_{\perp} \gtrsim \Gamma_e$ can be detected
\cite{honig66}, and optimal sensitivity is achieved at the threshold
for saturation, $\Omega_{\perp} = \Gamma_e$ \cite{desousa09}.  In
terms of the Kondo temperature, optimal EDMR takes place for
$\ln{(T/T_K)}=\sqrt{[3\pi/(2|p_e|)](B/B_{\perp})}$, a regime that requires
low Kondo temperature. We call this limit the ``weak coupling''
regime. For $B=1$~T and $B_{\perp}=0.3$~G, weak coupling requires
$\ln{(T/T_K)}\approx 500$.

We generalized Eqs.~(\ref{dots})--(\ref{dotsi}) for the case of
resonant excitation; when $\Omega_{\perp}\gtrsim\Gamma_e$ and
$\Omega_{\perp}\ll A$, the nuclear spin relaxation rates are given by
Eqs.~(\ref{iperp})--(\ref{iz}) with $\Gamma_e\rightarrow
\Omega_{\perp}$. Hence we have $T_{1n}\approx 2(\omega_e/A)^{2}
/\Omega_{\perp}$ and $T_{2n}\approx 2/\Omega_{\perp}$.  From this
result we can deduce the maximum possible contrast for read-out of
nuclear spin states: Nuclear spin-flips during read-out will lead to
contrast equal to
$\exp{(-T_{2n}/T_{1n})}=\exp{[-(A/\omega_e)^{2}]}$, that is
quite close to $1$ when $\omega_e\gg A$. We recall that this optimum
contrast occurs only when $t_{\rm{EDMR}}$, the time it takes to detect
EDMR of a single donor, is less than $T_{2n}$.

A more likely scenario is that $t_{\rm{EDMR}}$ will be quite long. 
Assuming the longest acceptable $t_{\rm{EDMR}}$ is equal to
$T_{1n}$, we now derive a sensitivity condition for EDMR detection
of nuclear spins. Such a condition will depend critically on the value
of the EDMR current amplitude $(\Delta I)_{\rm{EDMR}}$, that is given
by the current difference when only one donor is on and off resonance
with a microwave field.  A key observation is that $(\Delta
I)_{\rm{EDMR}}$ must be larger than the shot noise $(\Delta
I)_{\rm{shot}}$ accumulated during $T_{1n}$:
\begin{equation}
\left(\frac{\Delta I}{I}\right)_{\rm{EDMR}}> \left(\frac{\Delta I}{I}\right)_{\rm{shot}}=\frac{1}{\sqrt{N(T_{1n})}},
\label{ineq}
\end{equation}
where $I$ is the average device current and $N(T_{1n})=I T_{1n}/e$ is
the number of electrons that passed through the device during
$T_{1n}$. Using the value of $T_{1n}$ computed above we get the
sensitivity criterion for nuclear spin read-out:
\begin{equation}
\left(\frac{\Delta I}{I}\right)_{\rm{EDMR}} > \sqrt{\frac{e\Omega_{\perp}}{2I}}\left(\frac{A}{\omega_e}\right).
\label{criterion}
\end{equation}
We note that an implicit assumption of Eq.~(\ref{criterion}) is that
$A>1/T_{2e}^{*}$, i.e., the separation of the two EDMR peaks is larger
than their linewidth.  For $I=1$~$\mu$A, $B_{\perp}=0.3$~G and $B=1$~T
we get $(\Delta I/I)_{\rm{EDMR}} > 3\times 10^{-6}$ \cite{notenoise}.

A recent theory of the spin-dependent scattering mechanism of EDMR
\cite{desousa09} predicted $(\Delta I/I)_{\rm{EDMR}} = 6\times
10^{-6}\times \tanh{(\hbar\omega_e/2k_BT)}\times
(B/1\rm{T})\times (B_{\perp}/0.3\rm{G})$, for a single donor implanted in a
transistor of area $0.1\;(\mu \rm{m})^{2}$.  At $B\sim 1$~T and $T=1$~K this is
right at the sensitivity limit; however, much better sensitivity can
be obtained by going to higher $B$ fields.

There exists other methods for EDMR detection that can potentially
yield much higher $(\Delta I/I)_{\rm{EDMR}}$, most notably the ones
based on optical excitation and recombination
\cite{honig66,boehme03,mccamey06,stegner07,morley08}.  Electron-hole
pairs are excited by light and the photocurrent is monitored as a
function of magnetic field.  In this case additional relaxation
channels such as capture of an extra donor electron with rate
$\Gamma_c$, with corresponding ionization of the extra electron with
rate $\Gamma_i$ can be present. It is straightforward to generalize
our theory to these cases; we obtain the same expressions as above
with the substitution $\Gamma_e\rightarrow (\Gamma_e + \Gamma_i +
\Gamma_c)$ and $A^{2}\rightarrow \langle A^{2}\rangle =
A^{2}\Gamma_i/(\Gamma_i+\Gamma_c)$ (effective hyperfine coupling gets
reduced by capture, because a singlet produces zero hyperfine field on
the nuclear spin).  With these substitutions,
Eqs.~(\ref{1t2n})--(\ref{criterion}) can be applied to recombination
based EDMR.

In a typical transistor, the gate voltage can be used to control the
overlap between the electron gas and the donor impurity wavefunction.
Thus, the Kondo temperature can be controlled electrically from
$T_K=0$ (when the electron gas density is zero, i.e., the transistor
is turned off) all the way to large $T_K$. This is demonstrated for a
top gated device \cite{lo11,lo07} in Figure~\ref{fig1}.  Hence, we argue
that it is possible to switch donors electrically into and out of
three distinct modes: In the ``quantum evolution'' mode, donors have
no overlap with the electron gas, implying $T_K=0$ and $T_{1n},
T_{2n}$ are longer than a few seconds \cite{morton08}.  For larger gate
voltage, overlap is small and the donor is in the weak coupling
regime, allowing quantum read-out. For even larger gate voltages, the
donor will be in the strong coupling regime enabling fast initialization.

In conclusion, we presented a theory for the interaction of an
electric current with donor nuclear spin qubits. We showed that
electric tuning of donors away from the Kondo regime allows coherent
evolution and high contrast quantum read-out using electrically
detected magnetic resonance; in contrast, electric tuning close to the
Kondo regime enables fast initialization of nuclear spin qubits.  We
derived a general sensitivity criterion for the read-out of nuclear
spin states using EDMR.

We thank C.C. Lo, J.J.L. Morton, T. Schenkel, T. Tiedje, and L.H. Willems van
Beveren for useful suggestions to this work.  Our research was
supported by the NSERC Discovery program.


\begin{thebibliography}{99}
  
\bibitem{honig66} A. Honig, \prl {\bf 17}, 186 (1966); J. Schmidt
  and I. Solomon, C.R. Acad. Sci. B {\bf 263}, 169 (1966); D. Kaplan,
  I. Solomon, and N.F. Mott, J. Phys.  (Paris), Lett. {\bf 39}, L51
  (1978).

\bibitem{kane98} B.E. Kane, Nature {\bf 393}, 133 (1998).

\bibitem{boehme03} C. Boehme and K. Lips, \prl {\bf 91}, 246603 (2003).

\bibitem{mccamey06} D.R. McCamey, H. Huebl, M.S. Brandt, W.D.
  Hutchison, J.C. McCallum, R.G. Clark, and A.R. Hamilton, \apl {\bf
    89}, 182115 (2006).

\bibitem{stegner07} A.R. Stegner, C. Boehme, H. Huebl, M. Stutzmann, K. Lips, and M.S. Brandt, Nature Physics {\bf 2}, 835 (2007). 

\bibitem{sarovar08} M. Sarovar, K.C. Young, T. Schenkel, and K.B. 
  Whaley, \prb {\bf 78}, 245302 (2008).

\bibitem{morley08} G.W. Morley, D.R. McCamey, H.A. Seipel,
  L.-C. Brunel, J. van Tol, and C. Boehme, \prl {\bf 101}, 207602 (2008).

\bibitem{desousa09} R. de Sousa, C.C. Lo, J. Bokor, \prb  {\bf 80}, 045320 (2009).

\bibitem{morishita09} H. Morishita, L.S. Vlasenko, H. Tanaka, K.
  Semba, K. Sawano, Y. Shiraki, M. Eto, and K.M. Itoh, \prb {\bf 80},
  205206 (2009).

\bibitem{mccamey10} D.R. McCamey, J. Van Tol, G.W. Morley, and C. Boehme, Science {\bf 330}, 1652 (2010).

\bibitem{hoehne11} F. Hoehne, L. Dreher,H. Huebl, M. Stutzmann, and M.S. Brandt, \prl {\bf 106}, 187601 (2011).

\bibitem{lo11} C. C. Lo, V. Lang, R.E. George, J.J.L. Morton, A.M.
  Tyryshkin, S.A. Lyon, J. Bokor, and T. Schenkel, \prl {\bf 106},
  207601 (2011).

\bibitem{jelezko04} F. Jelezko, T. Gaebel, I. Popa, M. Domhan, A.
  Gruber, and J. Wrachtrup, \prl {\bf 93}, 130501 (2004); L.
  Childress, M. V. Gurudev Dutt, J. M. Taylor, A. S. Zibrov, F.
  Jelezko, J. Wrachtrup, P. R. Hemmer, and M. D.  Lukin, Science {\bf
    314}, 281 (2006).

\bibitem{ladd05} T.D. Ladd, D. Maryenko, and Y. Yamamoto, E. Abe, and
  K. M. Itoh, \prb {\bf 71}, 014401 (2005).

\bibitem{morton08} J.J.L. Morton, A.M. Tyryshkin, R.M. Brown, S.
  Shankar, B.W. Lovet, A. Ardavan, T. Schenkel, E.E. Haller, J.W.
  Ager, and S.A. Lyon, Nature {\bf 455}, 1085 (2008).

\bibitem{morello10} A. Morello {\it et al.}, Nature {\bf 467}, 687
  (2010).

\bibitem{hewson93} A. C. Hewson, {\it The Kondo Problem to Heavy Fermions}
  (Cambridge University Press, Cambridge, England, 1993).

\bibitem{vanhaesendonck87} C. Van Haesendonck, J. Vranken, and Y.
  Bruynseraede, \prl {\bf 58}, 1968 (1987).

\bibitem{lobaskin05} D. Lobaskin and S. Kehrein, \prb {\bf 71}, 193303
  (2005); F.B. Anders and A. Schiller, \prb {\bf 74}, 245113 (2006).

\bibitem{lansbergen10} G.P. Lansbergen, G.C. Tettamanzi, J. Verduijn,
  N. Collaert, S. Biesemans, M. Blaauboer, and S. Rogge, Nano Lett.
  {\bf 10}, 455 (2010).

\bibitem{rikitake05} This model can be formally derived within the
  Born-Markov approximation (second order in $J$), see Y. Rikitake and
  H. Imamura, \prb {\bf 72}, 033308 (2005).

\bibitem{mccamey09} D.R. McCamey, J. van Tol, G.W. Morley, and C. Boehme, \prl {\bf 102}, 027601 (2009).

\bibitem{yang09} A. Yang {\it et al.}, \prl {\bf 102}, 257401 (2009).

\bibitem{abragam61} See section IX-I-A of A. Abragam, {\it Principles
    of Nuclear Magnetism} (Oxford University Press, Oxford, England,
  1961).

\bibitem{lo07} C.C. Lo, J. Bokor, T. Schenkel, J. He, A.M. Tyryshkin,
  and S.A. Lyon, \apl {\bf 91}, 242106 (2007).

\bibitem{notenoise} We do not expect ``excess'' noise (due to other
  sources, e.g. $1/f$ noise) to be larger than shot noise during
  read-out.  In \cite{lo07}, EDMR was measured in a high quality
  transistor using lock-in detection at $1$~KHz; the excess noise
  after averaging for several minutes was $\Delta I/I\approx 10^{-8}$
  (see Fig.~2c in \cite{lo07}).  We remark that this is much lower
  than our shot noise estimate $(\Delta I/I)_{\rm{shot}} = 3\times
  10^{-6}$.


\end{thebibliography}
\end{document}